\documentclass[amsmath,amssymb,preprint,showpacs]{revtex4}

\usepackage{graphicx}
\usepackage[usenames]{color}

\begin{document}

\title{Evidence for coupling between collective state and phonons in
two-dimensional charge-density-wave systems}
\author{M. Lavagnini$^1$, M. Baldini$^2$, A. Sacchetti$^1$, D. Di Castro$^2$,
B. Delley$^3$, R. Monnier$^1$, J.-H. Chu$^4$, N. Ru$^4$, I.R. Fisher$^4$, P. Postorino$^2$ and
L. Degiorgi$^{1}$} \affiliation{$^1$Laboratorium f\"ur Festk\"orperphysik, ETH -
Z\"urich, CH-8093 Z\"urich, Switzerland} \affiliation{$^2$CNR-INFM-Coherentia
and Dipartimento di Fisica, Universit\`a ``La Sapienza'', P.le A. Moro 5,
I-00185 Rome, Italy} \affiliation{$^3$Paul Scherrer Institute, CH-5232 Villigen
PSI, Switzerland} \affiliation{$^4$Geballe Laboratory for Advanced Materials
and Department of Applied Physics, Stanford University, Stanford, California
94305-4045, U.S.A.}

\date{\today}

\begin{abstract}
We report on a Raman scattering investigation of the charge-density-wave (CDW), quasi two-dimensional rare-earth tri-tellurides $R$Te$_3$ ($R$= La, Ce, Pr, Nd, Sm, Gd and Dy) at ambient pressure, and of LaTe$_3$ and CeTe$_3$ under externally applied pressure. The observed phonon peaks can be ascribed to the Raman active modes for both the undistorted as well as the distorted lattice in the CDW state by means of a first principles calculation. The latter also predicts the Kohn anomaly in the phonon dispersion, driving the CDW transition. The integrated intensity of the two most prominent modes scales as a characteristic power of the CDW-gap amplitude upon compressing the lattice, which provides clear evidence for the tight coupling between the CDW condensate and the vibrational modes.
\end{abstract}



\maketitle

The electron-phonon coupling is of fundamental relevance for the development of several types of charge ordering in solids, of which the charge-density-wave (CDW) state, first predicted by Peierls \cite{peierls}, is an interesting realization. Peierls argued that one-dimensional (1D) metals are intrinsically unstable, and that a new broken-symmetry ground state results from the
selfconsistent rearrangement of the electronic charge density in response to
the (static) modulation of the ionic positions \cite{peierls,mazin}.
The new lattice periodicity leads moreover to the opening of a gap at the Fermi level. The consequences of this intimate connection between electronic properties and lattice dynamics 
have been intensively investigated in a number of prototype quasi-1D materials \cite{grunerbook}.

CDW's have been observed in transition metal di- and trichalcogenides \cite{wilsonADVPHYS,rouxel}, in the ladder compounds Sr$_{14-x}$Ca$_x$Cu$_{24}$O$_{41}$ \cite{vuletic} and in some copper oxide high temperature superconductors \cite{kivelsonRMP} (where they are known as ``stripes") as well, suggesting that similar effects are to be expected also in layered quasi-2D systems. A recent theoretical study \cite{kivelson} confirms that this is indeed the case, and that 
 two orthogonal CDW's may even combine to generate a checkerboard-like charge pattern. 
 However, high temperature superconductors are bad candidates for a systematic study of the interplay between electronic and phononic degrees of freedom in quasi-2D materials, given the strongly correlated nature of the electrons in these  systems. One class of quasi-2D compounds well suited to address this issue are the rare-earth ($R$) tri-tellurides \cite{dimasi,samples,ru}. They host a
CDW state already at 300 K and their structure consists of alternating
double Te ($ac$) planes (where the CDW resides) sandwiched between $R$Te layer blocks
and stacked along the long $b$ axis in the weakly orthorhombic (pseudo-tetragonal) cell. In our first
optical investigations, we have established the excitation across the CDW-gap
and discovered that this gap is progressively reduced upon compressing the
lattice either with chemical substitution (i.e., by changing $R$) or with
externally applied pressure \cite{sacchettiprb,sacchettiprl}. 


The formation of the CDW condensate in $R$Te$_3$ only partially gaps the Fermi surface \cite{brouet} and therefore these materials remain metallic even well below the
critical temperature $T_{CDW}$ at which the CDW appears \cite{ru}. This prevents the
investigation of the 
phonon modes and more generally of the impact
of the lattice dynamics on the CDW state \cite{mazin} in an infrared absorption experiment, as the corresponding signals are overwhelmed by the
metallic contribution. We therefore address the issue of the coupling
between vibrational modes and CDW condensate in these prototype 2D systems from the
perspective of the Raman scattering response. Our data, combined with the
measured CDW-gap, and supported by first principles calculations, allow us to
identify the expected Raman active modes and to determine their evolution under
pressure \cite{note4}. 
We provide clear cut evidence for a tight coupling between CDW
condensate and vibrational modes as well as robust predictions for the
incipient Kohn anomaly, i.e. the ``freezing in''  of a lattice  distortion
associated with the formation of the CDW phase \cite{peierls,grunerbook,mazin}.

Our  Raman scattering experiments were performed on the rare-earth series $R$Te$_3$ ($R$=La, Ce, Pr, Nd, Sm, Gd and Dy, with $T_{CDW}$ $>$ 300 K \cite{ru}) at ambient pressure as well as on LaTe$_3$ and CeTe$_3$ under externally applied pressure. The single crystalline samples of $R$Te$_3$ were grown by slow cooling of a binary melt \cite{samples}. Raman spectra were collected on cleaved [010] surfaces with a commercial micro-Raman spectrometer equipped with a He-Ne laser (632.8 nm wavelength, 16 mW power), a $20\times$ microscope objective (10~$\mu$m$^2$ laser spot), a notch filter to reject the elastic contribution, a 1800 lines/mm grating, 
and a cooled charge-coupled-device detector.
Additionally we have performed polarization-dependent Raman experiments on LaTe$_3$ at ambient pressure by varying the angle between the incident light polarization and the crystal axes by means of a $\lambda /2$ polarization rotator and selecting the scattered polarization parallel to the incident one. High pressures were generated by means of a Betsa membrane diamond-anvil-cell (DAC) equipped with high-quality type IIA diamonds (800~$\mu$m culet diameter) and a stainless steel gasket (300~$\mu$m hole diameter and 50~$\mu$m thickness). A small ($\approx 100\times 100$~$\mu$m$^2$) sample piece was placed inside the DAC together with the pressure transmitting medium (a 4:1 methanol-ethanol mixture \cite{metheth}) and a small ruby chip for pressure measurement  \cite{Mao}. 


\begin{figure}[!tb]
\center
\includegraphics[width=8.5cm]{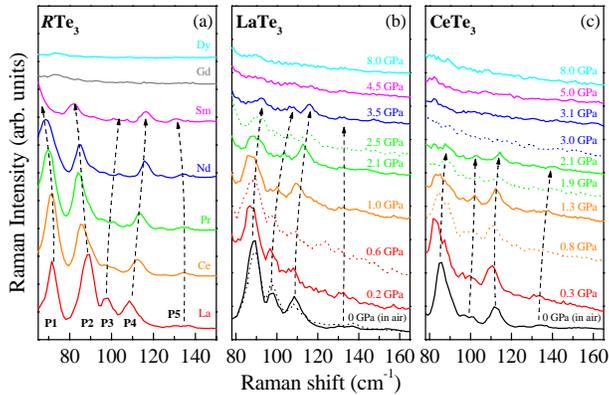}
\caption{(color online) Raman scattering spectra at 300 K for the $R$Te$_3$ series at ambient pressure (a), and for LaTe$_3$ (b) and CeTe$_3$ (c) for increasing
(continuous lines) and decreasing (dashed lines) pressure. All spectra have been
shifted for clarity.} \label{Experimental}
\end{figure}

Figure~\ref{Experimental}(a) summarizes the Raman scattering spectra, collected
for the whole $R$Te$_3$ series (i.e., chemical pressure). Four distinct modes
at 72, 88, 98 and 109 cm$^{-1}$ and a weak bump at 136 cm$^{-1}$ (labeled
P1-P5, respectively) can be identified in the La compound. The P1 mode slightly
softens from La to Nd and slowly moves outside the measurable spectral range at
ambient pressure (i.e., in SmTe$_3$ only its high frequency tail is still
observable). The remaining modes weakly disperse and progressively disappear
when going from the La to the Dy compound along the rare-earth series.

Panels (b) and (c) in Fig.~\ref{Experimental} display the Raman scattering
spectra of LaTe$_3$ and CeTe$_3$, under increasing and decreasing externally
applied pressure. 
The spectral range covered within the DAC is limited at low frequencies at
about 75 cm$^{-1}$, so that the lowest zero-pressure mode P1 cannot be clearly
detected in the applied pressure experiment. As in the chemical pressure case,
all other modes slightly disperse and disappear upon applying pressure. This
qualitative equivalence between chemical and applied pressure is also supported
by the fact that the peaks in LaTe$_3$ disappear at a slightly higher pressure
than in CeTe$_3$.
The pressure dependence is fully reversible since upon
decreasing pressure the modes reappear again.

\begin{figure}[!t]
\center
\includegraphics[width=8.5cm]{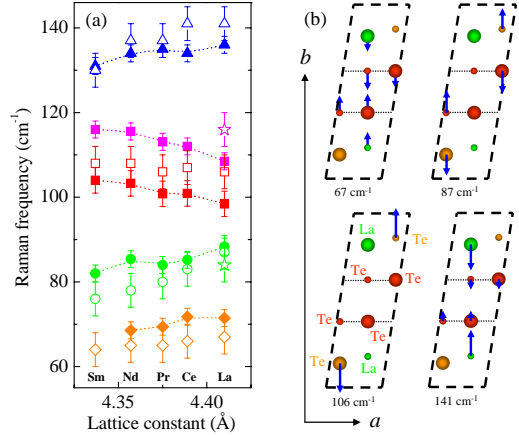}
\caption{(color online) (a) Raman-active phonon frequencies of $R$Te$_3$, obtained
experimentally (solid symbols) and calculated for the undistorted structure
(open symbols). The calculated extra peaks arising in the distorted structure (see text) are
also shown with star symbols for LaTe$_3$. The lattice constant values are from Ref. \onlinecite{Villars}. (b) Atomic displacements for the Raman
active $A_{1g}$ vibrational modes in LaTe$_3$. The figure shows the $ab$ plane, whereas large (small) spheres
represent atoms having a positive (negative) $c$ coordinate. The dotted lines represent the Te planes. The primitive
unit-cell and crystal axes are also shown. Arrows-lengths are proportional to
the calculated displacements.} \label{eigenvec-freq}
\end{figure}

The space-group of the undistorted structure is $Cmcm$ ($D_{2h}^{17}$) for all
rare earth tri-tellurides. From the occupied atomic positions \cite{struct} and
the factor-group analysis we determine the symmetry and multiplicity of the
Raman-active phonons \cite{GroupTh}, namely: $4A_{1g}+4B_{1g}+4B_{3g}$. The
corresponding Raman tensors imply that for our experimental configuration, in
which both incident and scattered light are polarized parallel to the $ac$
crystal plane, only the $A_{1g}$ symmetry phonons can be observed. 

The vibrational modes at the $\Gamma$ point of the Brillouin zone for the undistorted
structure 
at ambient pressure have
been obtained from first principles, using the Dmol$^{3}$ code developed by one
of us \cite{Delley}.  First, the positions of the 8 atoms in the primitive unit
cell were optimized at the experimental lattice constants $a$ and $b$ \cite{Tetragonal}, obtained
by averaging the values listed in the tables of Ref.
\onlinecite{Villars}. A frozen-phonon calculation then yielded the 24
sought-after frequencies. Only 4 of these correspond to the
expected Raman active modes with $A_{1g}$ symmetry.

\begin{figure}[!t]
\center
\includegraphics[width=8.5cm]{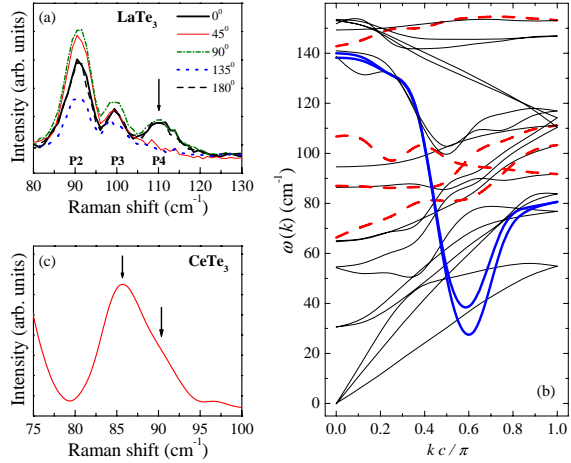}
\caption{(color online) (a) Raman spectra of LaTe$_3$ as a function of the polarization angle of
the incident light. (b) Calculated phonon dispersion along the $c$ axis. (c) Enlargement of the
P2 peak in CeTe$_3$ detailing its double-feature nature.}
\label{polariz-disp-2peaks}
\end{figure}

The frequencies of the calculated $A_{1g}$ modes at $\Gamma$ for the $R$Te$_3$ series are summarized in
Fig.~\ref{eigenvec-freq}a along with the experimental values, while Fig.~\ref{eigenvec-freq}b pictures the $A_{1g}$ lattice displacements of the undistorted structure, which, as predicted from the
factor-group analysis, are along the $b$ axis. The agreement with the experimental findings is
satisfactory. There is an obvious assignment of the calculated modes, for
instance at 67, 87 and 141 cm$^{-1}$ for LaTe$_3$, with the corresponding
features in the measured spectra (P1, P2, and P5). The calculated mode at about
106 cm$^{-1}$ lies between the experimentally observed modes P3 and P4 for
LaTe$_3$. This situation persists throughout the whole rare earth series
(Fig.~\ref{eigenvec-freq}a). Our polarization dependent measurements on
LaTe$_3$ (Fig.~\ref{polariz-disp-2peaks}a) yield an angle-dependent intensity
with a period of $90^{\circ}$ for the P4 mode and $180^{\circ}$ for the adjacent peaks P2 and P3 \cite{note3}. Since a $180^{\circ}$ period is expected for the $A_{1g}$
symmetry, the P4 mode cannot be assigned
within the undistorted structure. 

In order to improve the mode assignment and to clarify the origin of the P4 peak, we have, in a
first step, computed the phonon dispersion in LaTe$_3$ along the $\Gamma - Z$
direction of the Brillouin zone, using a supercell consisting of a 16-fold repetition along
the $c$-axis of the primitive unit cell. This yields phonon frequencies at wave
vectors $(0,0,q_{z})$, where $q_{z}=\frac{n}{16}c^*$, with $c^*=\frac{2\pi}{c}$,
and $n$ between 0 and 8. A symmetry constrained spline interpolation between
these results is shown in Fig.~\ref{polariz-disp-2peaks}b (similar
results have been obtained along the $a$ axis). The dashed branches
highlight the phonon dispersion for the modes with $A_{1g}$ symmetry at
$\Gamma$. The calculated phonon spectrum also shows a distinct
Kohn anomaly at $q_{z}$ slightly below $0.3c^*$ (thick lines), i.e. in the region
expected from the electron diffraction results 
\cite{dimasi}
and the angle-resolved photoemission spectroscopy data on CeTe$_{3}$ 
\cite{brouet}. 
In
a second step, we have generated a commensurate approximant to the distorted
structure in the presence of the incommensurate CDW by repeating the calculation for a 14-fold repetition of the primitive unit cell along the $c$-axis (in which case a true instability
occurs at $q_{z}=\frac{2}{7}c^*$, i.e. the frequency of the soft phonon becomes
imaginary), then moving the atoms along the eigenvectors  of the soft phonon,
and reequilibrating their positions in the corresponding ($1 \times 1 \times
7$) supercell \cite{dist_struct}. As a consequence of the lower symmetry,
vibrational modes with $A_1$ and $B_1$ symmetry become Raman-active
\cite{Phon_Symm}. Although there are 56 $A_1$ symmetry modes, their frequencies
accumulate around those of the $A_{1g}$ modes of the undistorted structure,
suggesting that the distortion does not particularly affect the $\Gamma$-point
vibrational energies. This is consistent with the fact that the phonon branches having
$A_{1g}$ symmetry at $\Gamma$ are weakly dispersing
(Fig.~\ref{polariz-disp-2peaks}b). The main effect of the distortion is the
appearance of 28 $B_1$ modes, which accumulate around 84 and 116~cm$^{-1}$. The
latter frequency compares very nicely with the frequency of the P4 peaks in our experiment. The second $B_1$ mode at about 84 cm$^{-1}$ falls in the range of the P2 peak. A closer look at the experimental data (Fig.~\ref{polariz-disp-2peaks}c for CeTe$_3$) indeed suggests that the P2 peak
may be a double feature.

\begin{figure}[!t]
\center
\includegraphics[width=8.5cm]{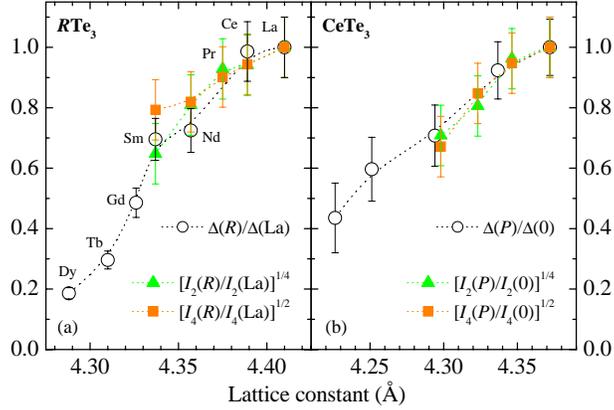}
\caption{(color online) Comparison between the amplitude of the CDW-gap (open circles)
\cite{sacchettiprb,sacchettiprl} and the integrated intensities ($I$) of the P2 (full triangles)
and P4 (full squares) peaks, raised to 1/4 and 1/2, respectively (see text)
as a function of the lattice constant for the $R$Te$_3$ series (a) and for
CeTe$_3$ under pressure (b). The pressure dependence of the lattice constant in CeTe$_3$ was achieved following the procedure of Ref. \onlinecite{sacchettiprl}. All quantities are normalized to their
starting values.} \label{Delta_Intensity}
\end{figure}

The main result of our experimental investigation is the observation of a systematic 
decrease of the integrated intensity ($I$) of the most prominent peaks P2 and P4 in the Raman 
spectra of Fig.~\ref{Experimental} with pressure \cite{note1}, which bears a striking similarity with the behaviour of the amplitude of the CDW-gap $\Delta$ (i.e., the order parameter) upon compressing the lattice, as obtained from the optical conductivity \cite{sacchettiprb,sacchettiprl}. One could first argue that these modes disappear because of an enhancement of their width and a concomitant decrease of their apparent amplitude due to the increase in free carrier concentration upon compressing the lattice \cite{sacchettiprb}. The modes' width remains, however, almost constant so that this possibility is rather unlikely. Our optical data \cite{sacchettiprb} also allow us to exclude the possibility that the phonon modes disappear due to an increase of the absorption coefficient at the laser frequency with decreasing lattice constant. Figure~\ref{Delta_Intensity} shows that the intensities of the P2 and P4 peaks  scale 
with $\Delta^4$ and $\Delta^2$, respectively \cite{note2}, suggestive of a coupling between the lattice vibrational modes and the CDW condensate. This is not at all surprising for the P4 mode, as
our calculations predict this peak only in the distorted structure. For the P2
peak we should consider its two components, namely the $A_{1g}$ mode in the
undistorted structure and the $B_1$ mode in the distorted one (Fig. 3c). For the latter the
intensity is obviously correlated with the CDW, whereas for the former at 87 cm$^{-1}$ the
correlation can be explained by looking at the corresponding atomic
displacements (Fig.~\ref{eigenvec-freq}b), which strongly distort the Te-planes
and therefore should couple to the CDW \cite{rice}. Furthermore, the specific behaviour ($I\sim\Delta^q$, $q$=2 or 4) is consistent with theoretical
predictions for the intensity in the distorted phase of originally silent modes,
obtained from a group theoretical analysis in the framework of Landau's
theory of second order phase transitions \cite{Phon_Intensity}.

In summary, we have been able to draw a consistent picture of the Raman
response of the CDW rare-earth tri-tellurides, by combining experimental
observations and numerical simulations. In particular, 
we have provided clear evidence for the tight
coupling between the CDW-gap and the lattice degrees of freedom
and have made a robust prediction for the Kohn anomaly inducing the CDW phase transition. Therefore, the easily tunable $R$Te$_3$ series provides a rather unique playground for a systematic study of the mechanism leading to the formation of the CDW state.\\

The authors wish to thank S.L. Cooper, Z.X. Shen and R. Hackl for fruitful discussions.
This work has been supported by the Swiss National Foundation for the Scientific Research and by the NCCR MaNEP pool, as well as by
the Department of Energy, Office of Basic Energy Sciences under contract
DE-AC02-76SF00515.


\begin{thebibliography}{99}
\bibitem{note4} As the temperature dependence of the CDW-gap is not known in $R$Te$_3$, we choose pressure (both chemical and applied) as tunable variable.
\bibitem{note3} We also collected data (not shown here) with the scattered light polarization orthogonal to the incident one and obtained results fully consistent with those of the parallel polarization (Fig. 3a).
\bibitem{Phon_Symm} The unit cell in the distorted structure is not centrosymmetric anymore, so that 
the $A_{1g}$ symmetry turns into $A_1$. The period of the angle-dependent intensity is $180^{\circ}$ ($90^{\circ}$) for the $A_1$ ($B_1$) symmetry.
\bibitem{note1} The sudden disappearance of the modes in the Raman spectra of Gd and Dy tri-tellurides (Fig. 1a) could also originate from the fact that the data were collected at 300 K, i.e. just slightly below
their own $T_{CDW}$ \cite{ru}.
\bibitem{note2} The behaviour of the weak 
P3 and P5 peaks is less well defined, while the intensity of the P1 peak, which is 
only partially seen in the chemical-pressure series, does not seem to change much.
\end{thebibliography}
\end{document}